\documentstyle[onecolumn]{mn}
\input mpp.sty
\input epsf.sty

\begin{document}
\title[Propagation effect on polarization of pulsar]{Propagation effect on polarization of
            pulsar radio emission}
\author[ R. T. Gangadhara, H. Lesch, V. Krishan]
{R. T. Gangadhara$^{1,3}$, 
H. Lesch$^{^2}$ and V. Krishan$^1$\\
$^1$ Indian Institute of Astrophysics, Bangalore--560034, India\\
$^2$Institut f\"ur Astronomie und Astrophysik der Universit\"at M\"unchen,
Scheinerstrasse 1, 81679 M\"unchen, Germany\\
$^3$National Centre for Radio Astrophysics, TIFR, Pune University Campus, Pune--411007, India\\
}
\date{Received:}
\maketitle
\begin{abstract}
We consider the role of a propagation effect such as the
stimulated Raman scattering on the polarization of radio pulses. When
an intense electromagnetic wave with frequency close to the plasma frequency
interacts with the plasma in the pulsar magnetosphere, the incident wave
undergoes stimulated Raman scattering. Using typical plasma and magnetic field
parameters, we compute the growth rate and estimate the polarization
properties of the scattered mode. At some conditions, we find that the polarization
properties of the
scattered mode can become significantly different from those of the incident
wave. The frequencies, at which strong Raman scattering occurs
in the outer parts of magnetosphere, fall into the observed
radio band.
 In some pulsars, for example, PSR B0628-28 and 1914+13, a
large and symmetric type of circular polarization has been observed. We
propose that such an unusual circular
polarization is produced by the propagation effects.
\end{abstract}

\begin{keywords}
 stars:pulsars-plasmas-waves-radiative transfer
\end{keywords}

\section{Introduction}
   The investigation of coherent radio emission from the pulsar
magnetosphere has attracted a great deal of attention (Cordes 1979; Michel 1982;
Ass\'eo et~al. 1990).
A powerful collective emission occurs when relativistic electron beams with
density $\sim$ 1 per cent of the pair plasma density, scatter off
coherently from concentrations of plasma waves (cavitons) (Benford 1992).
 The role of collective plasma processes in the absorption and spectral
modification of the radio waves is well known (e.g., Beal 1990; Krishan \& Wiita
1990; Benford 1992; Gangadhara \& Krishan 1992, 1993, 1995;
Gangadhara,  Krishan \& Shukla, 1993). von~Hoensbroech, Lesch and Kunzl (1998)
have demonstrated that degree of linear polarization decreases with increasing
frequency while the degree of circular polarization shows the opposite trend.
\par
In this paper, we estimate the role of stimulated Raman scattering
on the polarization of electromagnetic waves propagating in the pulsar magnetosphere.
We assume that the physical conditions in the pulsar magnetosphere are those of the
classical standard model (Ruderman \& Sutherland 1975) which describes the
generation of ultrarelatvistic beams of electrons/positrons and the creation
of the pair plasma. The beams and the pair plasma are in relativistic motion along
the bundle of open magnetic field lines that delimit the active region of
magnetosphere.
The stimulated Raman scattering is considered as a parametric decay of the
initial transverse electromagnetic (pump) wave into another electromagnetic wave and
a longitudinal plasma wave. The physics of stimulated Raman scattering in a plasma
has been explained in many papers and books (e.g.,
Drake et al. 1974; Liu \& Kaw 1976; Hasegawa 1978; Kruer 1988).
In two ways, stimulated Raman scattering may be important in the pulsar environment:
first, it may act as an effective damping
mechanism for the electromagnetic waves generated
by some emission mechanism at the lower altitudes in the
pulsar magnetosphere. At those altitudes, the resonant conditions for stimulated Raman
 scattering, i.e., the frequency and wave number matching might not  be satisfied.
This results in a short time variability which is generally
observed in pulsar radio emission. Secondly,
it may provide an effective saturation mechanism for the
growth of the electromagnetic waves provided  that the conditions
for the wave excitation by some mechanism are satisfied
in the region where an effective stimulated Raman scattering can take place.
\par
The first case can be simplified by treating the intensity of the pump
is constant in time. Then the nonlinear equations, which describe the wave coupling,
become linear in amplitudes of the decaying waves, and the exponentially growing
solutions will imply an effective energy transfer from the pump wave.
This approximation breaks down when the amplitudes of the decay waves become
comparable to the pump wave or when the amplitudes of the decay waves enter nonlinear
stage, and start loosing energy due to some nonlinear processes such
as wave-particle trapping and acceleration.
\par
The second case is more complicated, where stimulated Raman scattering acts as a nonlinear saturation
mechanism and the amplitudes of all waves may be of the same order. This case can be
considerably simplified when damping of the plasma wave is very strong or if it
leaves the region of the resonant interaction fast enough.
\par
We neglect the nonlinear stages of stimulated Raman scattering and the
development of Langmuir turbulence, which leads to wave--particle trapping or
quasilinear diffusion. If the pump is monochromatic, the growth rate of stimulated
Raman scattering can become very high, as in conventional laboratory laser-plasma
interaction. However, in the case of pulsars the pump can be broadband, and
in the limit where the bandwidth $\Delta \omega$ of the pump wave
is much larger than the growth rate of stimulated Raman scattering, we can use a random phase approximation
for the statistical description of the interacting waves.
\par
 Tsytovich and Shvartsburg (1966) have given a general expression for the third order
nonlinear currents excited in a magnetized plasma. Since the corresponding
expressions  are very complicated, the general case of Raman scattering becomes
very difficult to consider. However, one can make some useful simplifications, when
considering stimulated Raman scattering in the pulsar magnetosphere. First, in
superstrong magnetic field,
we can expand the currents in $1/\omega_B,$ where $\o_B$ is the cyclotron frequency.
Second, if the  pair plasma has the same distributions for electrons and positrons then
some of the  nonlinear currents cancel out, as they are proportional to the third power
of the electric charge (this cancellation is exact in the unmagnetized electron-positron
plasma). Third, all the three interacting waves propagate along magnetic field.
This is an important but less justified approximation. It allows us to simplify the
problem considerably, and to obtain a dispersion relation for stimulated Raman 
scattering.
\par
The polarization of the pulsar signals appear to bear critically on the
pulsar radio emission process and the emission beam geometry. One or more reversals
of the sense of circular polarization
has been observed in the intergrated profiles of several pulsars. However,
in individual pulses circular polarization changes sense many times across
the pulse window. Further, it is important to determine whether the depolarization
is a geometric effect or results from radiation--plasma interactions.
There have  only been very preliminary attempts to explain
depolarization and microvariability using plasma mechanisms (Benford 1992).
\par
 Our purpose in this paper is to derive the properties of natural plasma modes
and to explore some possible implications concerning the interpretation of the
observed polarization, notably
large and symmetric circular polarization in some pulsars (e.g., PSR B0628-28,
and 1914+13). In \S 2, we derive the dispersion relation for stimulated Raman
scattering of an electromagnetic wave, and give an analytical expression for the
growth rate of the instability. In \S 3, we define the Stokes
parameters and compute the polarization states of the scattered electromagnetic
wave. The discussion of our findings is given in \S 4.

\section{Polarization changes due to stimulated Raman scattering}

We begin with a model consisting of a pulsar with  nonthermal
component of radiation interacting with plasma in the emission region
at a distance $r=100R_{NS}=10^{8}$~cm, (neutron star radius
$R_{NS}\approx 10$ km) from the neutron star surface, where the magnetic field is
about $10^6$~G. Plasma particles may be all in their lowest Landau level with
no Larmor gyration, however, plasma can have one-dimensional distributions of
momenta along the magnetic field (Blandford 1975; Lominadze, Machabelli
\& Usov 1983).
This is because the synchrotron loss time for the decay of the perpendicular
component of momentum is always short compared to the transit time at the
stellar surface for any velocity.
The simplest model for the plasma is density declining in proportion to
$r^{-3}$ with no gradients in the distribution functions across $\vec B.$
\par
The nonlinear interaction of radiation with plasma causes the modulation
instability
leading to enhancement of nonresonant density perturbations and the radiation
amplification by free-electron-maser, which produces intense electromagnetic
waves (Lesch, Gil \& Shukla 1994; Gangadhara, Krishan \& Shukla 1993;
Gangadhara \& Krishan 1992). Since the frequency of these electromagnetic waves
is close to the plasma frequency, they resonantly couple with the
subpulse-associated plasma column in the pulsar magnetosphere and
drive the stimulated Raman scattering.
\par
Consider a large amplitude electromagnetic wave
$(\vec k_i, \o_i)$ with an electric field:
\beq
\vec E_i=\eps_{xi}\cos(\vec k_i\cdot\vec r-\o_it)
\hat{e}_x+\eps_{yi}\cos(\vec k_i\cdot\vec r-\o_it+\d_i)\hat{e}_y
\eeq
interacts with the plasma in the pulsar magnetosphere.
\par
We follow Ruderman \& Sutherland's (1975) approach to estimate the
density and plasma frequency of the plasma moving within the bundle
of field lines. For the typical parameters:
Lorentz factor of primary particles $\g_p\sim 10^6,$ and for pair plasma particles
$\g_\pm\sim 10^3$, magnetic field $B_o\sim 10^6 r_8^{-3}$~G, and pulsar period
$p=1$~s, we obtain the particle number density
\beq
n_o=\frac{\g_p}{2\g_\pm}\frac{B_o}{e c p}
=\frac{3.5\times 10^7}{r_8^3}\quad{\rm cm}^{-3}
\eeq
and the plasma frequency
\beq
\o_{p}=2\g_\pm\bc(\frac{4 \pi n_o e^2}{\g_\pm m_o}\bc)^{1/2}
=\frac{2\times 10^{10}}{r_8^{3/2}}\quad\hbox{{\rm rad s}}^{-1}.
\eeq

The plasma in the emission region of pulsar magnetosphere may be birefringent
(Melrose \& Stoneham 1977; Melrose 1979; Barnard \& Arons 1986;
von Hoensbroech, Lesch and Kunzl 1998).
In these models, two modes of wave propagation is generally allowed
in magnetoactive plasma: one mode is polarized in the $\vec k_i$--$\vec B$
plane and other mode in the direction perpendicular to it.
Following these models, we resolve $\vec E_i$ into two modes
$\vec E_{xi}=\eps_{xi}\cos(\vec k_i\cdot\vec r-\o_it)\hat{e_x}$
and $\vec E_{yi}=\eps_{yi}\cos(\vec k_i\cdot\vec
r-\o_it+\d_i)\hat{e_y}$ such that they are polarized in the
directions parallel and perpendicular to the $\vec k_i$--$\vec B$ plane,
respectively. It is well known from laser-plasma interactions that
large amplitude electromagnetic waves resonantly interact with the
plasma when the radiation frequency becomes close to the plasma
frequency. Since the two modes have different indexes of refraction
(McKinnon 1997), i.e., the response of plasma is not same for the two
radiation modes, it is reasonable to assume that $\vec E_{xi}$
couples with the density perturbation $\D n_{1}=\d n_1 \cos(\vec k\cdot\vec r
-\o t),$ and $\vec E_{yi}$ couples with $\D n_{2}=\d n_2 \cos(\vec k\cdot\vec r
-\o t+\d)$ in the plasma medium.
\par
Since the ponderomotive force is proportional to $\n E_{i1}^2$ and
$\n E_{i2}^2$, the coupling between the radiation and the density
perturbations is nonlinear. Hence the density perturbations grow up which
lead to currents and mixed electromagnetic--electrostatic side-band
modes at ($\vec k_i\pm \vec k,\o_i\pm \o).$ In turn these side-band modes
couple with the incident wave field, producing a much stronger
ponderomotive force, which amplifies the original density perturbation.
Hence a positive feedback system sets in, which leads to an instability.
\par
The electric field $\vec E_s$ of the electromagnetic wave scattered through 
an angle
$\phi_s$ with respect to $\vec k_i$ can be written as
\beq
\vec E_s=\eps_{xs}\cos(\vec k_s\cdot\vec r-\o_st)\hat{e}'_x+\eps_{ys}
\cos(\vec k_s\cdot\vec r-\o_st +\d_s)\hat{e}'_y.
\eeq
\par
	The propagation directions of incident wave $(\vec k_i,\o_i)$ and
scattered wave $(\vec k_s, \o_s)$ are illustrated in Fig.~1, such that
$\vec k_i\parallel\hat{e}_z,$ $\vec k_s\parallel\hat{e}'_z$ and
$\hat{e}'_y\parallel \hat{e}_y.$ The primed coordinate system
is rotated through an angle $\phi_s$ about the y-axis. Then
the scattered wave in the unprimed coordinate system is given by
\beq
\vec E_s=\eps_{xs} \cos(\phi_s) \cos(\vec k_s\cdot\vec r-\o_st) \hat{e}_x
+\eps_{ys} \cos(\vec k_s\cdot\vec r-\o_st +\d_s)\hat{e}_y
-\eps_{zs}\sin(\phi_s) \cos(\vec k_s\cdot\vec r-\o_st)\hat{e}_z,
\eeq
where $\eps_{zs}=\eps_{xs}.$
\par
The quiver velocity $\vec u_\pm$ of positrons and electrons due to the
radiation fields $\vec E_i$ and $\vec E_s$ is given by
\beq
{{\partial\vec u_\pm}\over
{\partial t}}=\pm{e\over m_{0}}(\vec E_i+\vec E_s),
\eeq
 where $e$ and $m_{0}$ are the charge and rest mass of the plasma particle.
\begin{figure}
\vskip -2.5 cm
\hskip 2.5 cm
\epsfysize15.0 truecm {\epsffile[18 176 558 716]{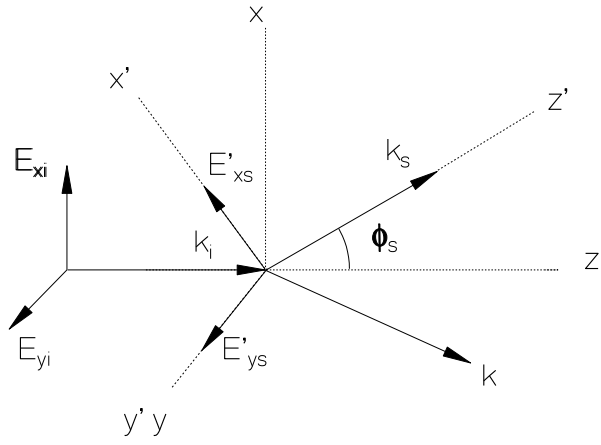}}
\vskip -7 cm
\caption{Stimulated Raman scattering of transverse electromagnetic wave
$(\vec E_{xi},\, \vec E_{yi})$ through an angle $\phi_s.$ The scattered wave
electric field is $(\vec E_{xs},\, \vec E_{ys}).$ The wave numbers $\vec k_i,$
$\vec k_s$ and $\vec k$ are due to the incident, scattered and Langmuir waves,
respectively.}
\end{figure}
\par
The wave equation for the  scattered electromagnetic wave is given by
\beq
\bc ( \n^2-{{1}\over {c^2}}{{\partial^2}\over
{\partial t^2}}\bc )\vec E_s={{4\pi}\over{c^2}}{{\partial\vec J}\over
{\partial t}},
\eeq
where $c$ is the velocity of light and $\vec J$ is the current density.

The components of equation~(7) are
\beq
D_s\eps_{xs}\cos(\phi_s)\cos(\vec k_s \cdot\vec r-\o_s t)
= -\frac{2\pi e^2}{m_{0}} \frac{\o_s}{\o_i} \eps_{xi}\d n_1
\cos(\vec k_s \cdot\vec
r-\o_s t)
\eeq
and
\beq
D_s \eps_{ys} \cos(\vec k_s \cdot\vec r-\o_s t+\d_s)
  = -\frac{2\pi e^2} {m_{0}} \frac{\o_s }{\o_i}
  \eps_{yi}\d n_2 \cos(\vec k_s \cdot\vec r-\o_s t+\d_i-\d)
\eeq
and
\beq
D_s \eps_{zs}\sin(\phi_s )\cos(\vec k_s \cdot\vec r-\o_s t) =0,
\eeq
where $D_s =k_s ^2c^2-\o_s^2+\o_{p}^2$ and
\beq
\o_s=\o_i-\o,\quad \quad \vec k_s=\vec k_i-\vec k.
\eeq
 In quantum language these two relations may be interpreted as the
conservation of energy and the momentum along the magnetic field,
respectively. When these conditions are satisfied, stimulated Raman scattering is exicted
resonantly and the expression $D_s \approx 0$ becomes the dispersion relation
of the scattered electromagnetic mode.

If we cancel the instantaneous space and time dependent cosine
functions on both sides of equation~(8), we get
\beq
	D_s\eps_{xs}\cos\phi_s =-\frac{2\pi e^2}{m_{0}}\frac{\o_s}{\o_i}
	\eps_{xi}\d n_1.
\eeq
Similar to the phase matching conditions (equation~11), equation~(9) gives the
condition between the initial phases:
  \beq
  \d_s=\d_i-\d,
  \eeq
and hence we obtain
\beq
D_s\eps_{ys}=-\frac{2\pi e^2}{m_{0}}\frac{\o_s}{\o_i}\eps_{yi}\d n_2.
\eeq
Dividing  equation~(14)  by equation~(12),  we have
\beq
\a_s=\a_i \eta\cos(\phi_s),
\eeq
where
$\a_i=\eps_{yi}/\eps_{xi},$ $\a_s=\eps_{ys}/\eps_{xs}$ and
$\eta={\d n_2}/{\d n_1}.$ The value of $\eta$ is determined by $\vec E_i$ in such
way that $\D n_{1}$
couples with $\vec E_{xi}$ and $\D n_{2}$ couples with $\vec E_{yi}$.
\par
We consider the Vlasov equation to find the low frequency plasma response:
\beq
{{\partial f}\over{\partial t}}+\vec v.\n f+{1\over m_0}
(e\n \phi-\n \psi).{{\partial f}\over{\partial\vec v}}=0,
\eeq
 where $\phi (\vec r, t)$ is the scalar potential associated with the
electrostatic waves, $f(\vec r,\vec v,t) $ is the particle distribution
function and $\psi (\vec r ,t)$ is the ponderomotive potential.
\par
Using $f(\vec r,\vec v,t)=f_0(\vec v)+ \D f_{1}(\vec r,\vec v,t)+
\D f_{2}(\vec r,\vec v,t),$  we can linearize equation~(16), and obtain:
\beq
{{\partial (\D f_{1}})\over{\partial t}}+{{\partial (\D
f_{2})}\over{\partial t}}+ \vec v.\n (\D f_{1})
+\vec v.\n (\D f_{2})+{{1}\over {m_0}}
( e\n \phi-\n\psi).{{\partial f_0}\over{\partial\vec v}}=0,
\eeq
where $\D f_{1}=\d f_1 \cos(\vec k\cdot\vec r-\o t)$ and $\D f_{2}=\d f_2
\cos(\vec k\cdot\vec r-\o t+\d).$
The ponderomotive force of the radiation field depends quadratically on the
amplitude of the electric field. Physically, it is a radiation pressure which
amplifies the density perturbations by exciting the slow longitudinal fields
and motions. The ponderomotive potential is given by:
\begin{eqnarray}
 \psi & = & {{e^2}\over{2 m_0}}\bc <
\bc (  \hbox{Re} \bc \vert\frac{\vec E_i}{i\o_i}+\frac{\vec E_s}{i\o_s}
\bc \vert\bc )^2\bc >_{\o} \nonumber \\
& = & {{e^2}\over{2 m_0\o_i^2}}[\cos(\phi_s)\cos(\vec k\cdot\vec r-\o t)+
\a_i\a_s \cos(\vec k\cdot\vec r-\o t+\d_i-\d_s)]\eps_i\eps_s,
\end{eqnarray}
  where the bracket $\bigl <\, \bigr >_{\o}$ represents the average over the
fast time scale $(\o_i\gg \o).$
\par
   In the presence of a strong magnetic field, the pair plasma becomes polarized
and one dimensional. If there is some relativistic drift between the electrons
and the positrons then the ponderomotive force becomes effective, which will induce
the nonlinear density perturbations (Ass\'eo 1993):
\beq
\d n_\pm \approx -\frac{1}{32\pi}\frac{{\vert\eps_\parallel\vert}^2}{k_B T_p}\frac{\o_p^2}
{\o_i^2}\frac{1}{\g_\pm},
\eeq
where $\eps_\parallel$ is the envelope of the parallel electric field of Langmuir wave
which is slowly varying with space and time, and $T_p$ is the plasma temperature.
Using these density perturbations and Poisson equation, we can self--consistently
derive the potential $\phi$:

\beq
\phi=-\frac{4\pi e}{k^2}(\D n_{1}+\D n_{2}).
\eeq
Now, substituting the expressions for $\D f_{1},$ $\D f_{2},$ $\D n_{1},$
$\D n_{2},$  $\phi$ and $\psi$ into equation~(17) and using the condition (13),
we obtain
\beq
\d f_2  +\d f_1\mu   +\bc [
\frac{4\pi e^2}{m_0 k^2} (\d n_2+\d n_1\mu) + \frac{e^2}{2 m_0^2 \o_i^2}
\{\cos(\phi_s) \mu+ \a_i\a_s\}\eps_i\eps_s
\bc ] {{\vec k.{{\partial f_0}\over
{\partial\vec v}}}\over {(\o-\vec k\cdot\vec v)}}=0,
\eeq
where $\mu={\sin(\vec k\cdot\vec r-\o t)}/{\sin(\vec k\cdot\vec r-\o t+\d)}.$
For $\d=0$ and $\pi,$  $\mu=\pm 1,$ while at the other valuse of $\d$
we have to find the average $<\mu >=\mu_a$ over the time scale $T=2\pi/\o.$ 
Therefore, we have
\beq
\d f_2+\d f_1 \mu_a=-\frac{4\pi e^2}{m_0 k^2}\bc [ \d n_2+\d n_1\mu_a
+\frac{\eps_i k^2}{8\pi m_0 \o_i^2} A \bc ] {{\vec k.{{\partial f_0}
\over{\partial\vec v}}}\over {(\o-\vec k\cdot\vec v)}},
\eeq
where $A=(\mu_a\cos\phi_s+ \a_i\a_s)\eps_s.$
The sum of density perturbations $(\d n_2+\d n_1\mu_a)$ can be estimated as
\beq
\d n_2+\d n_1\mu_a
=\int\limits_{-\infty}^{\infty} n_0 (\d f_2+\d f_1\mu_a) d\vec v
= -\bc [ \d n_2+\d n_1\mu_a
+\frac{\eps_i k^2}{8\pi m_0 \o_i^2}A \bc ] \chi,
\eeq
where
\beq
\chi=\frac{\o_{p}^2}{k^2}\int\limits_{-\infty}^{\infty}
{{\vec k.{{\partial f_0}\over{\partial\vec v}}}\over
{(\o-\vec k\cdot\vec v)}}d\vec v,
\eeq
is the susceptibility function (Liu \& Kaw 1976; Fried \& Conte 1961).
Using $\eta=\d n_2/\d n_1,$ we can write
\beq
\bc ( 1+\frac{1}{\chi}\bc) (\eta+\mu_a) \d n_1  =- \frac{\eps_i k^2}{8\pi m_0
\o_i^2} A.
\eeq
Multiplying equation (12) by $\eps_i\mu_a$ and equation (14) by $\a_i\eps_i$, 
and adding we get
\beq
(\a_s\a_i+\mu_a\cos\phi_s)\eps_s=-\frac{2\pi e^2}{m_o}\frac{\o_s}{\o_i}
\eps_i(\mu_a+\a_i^2\eta)\frac{\d n_1}{D_s},
\eeq
where $\eps_i=\eps_{xi}$ and $\eps_s=\eps_{xs}.$

Now, using equations~(25) and (26), we obtain the dispersion relation for stimulated Raman scattering:
\beq
\bc (1+\frac{1}{\chi}\bc) (\eta+\mu_a) =\frac{v_0^2k^2}{4}
\frac{(\mu_a+\a_i^2 \eta)}{{(1+\a_i^2)}}\frac{\o_s}{\o_i}\frac{1}{D_s},
\eeq
where $v_0=e\eps_i\sqrt{1+\a_i^2}/m_0\o_i$ is the
quiver velocity of plasma particles due to the electric field of incident
electromagnetic wave.
\par
If $L=L_{30}\times 10^{30}$ erg s$^{-1}$
is the luminosity of pump radiation with frequency $\o_i=2\pi\nu_i=\nu_{10}
\times 10^{10}$~rad~s$^{-1}$ at a distance $r=r_8\times 10^8$~cm from the source then
\beq
v_0 =\frac{e}{m_0\o_i}\bc (\frac{2L}{r^2 c}\bc)^{1/2}
=4.3\times 10^9\frac{L_{30}^{1/2}}{\nu_{10} r_8}\quad{\rm cm~s}^{-1}.
\eeq
When the phase matching conditions are met, the instability becomes more
efficient, and
$D_s=2\o_i\o-\o^2-2c^2\vec k_i.\vec k+c^2k^2\approx 0$ represent the dispersion
relation of the Stokes mode.
\par
The thermal speed $v_t$ of the plasma can be expressed in terms of
its energy spread in the laboratory frame. Let
$v_\pm=c(1-1/\g_\pm^2)^{1/2}$ be the velocity of electron-positron plasma, then
the velocity spread $\d v_\pm$ is given by (Hasegawa 1978; Gangadhara \& Krishan
1992)
\beq
\d v_\pm\approx c\frac{\d \g_\pm}{\g^3} \quad \quad {\rm for}\quad \g\gg 1.
\eeq
Now, using the Lorentz transformation of velocity, we can show that
\beq
\d v_\pm=\d v_z=\frac{\d v_z'}{\g_\pm^2(1+v_\pm v_z'/c^2)}
\approx \frac{\d v_z'}{\g^2}=\frac{v_t}{\g^2}
\eeq
because $v_z'=0.$ Hence the thermal speed, in the plasma frame, is given by
\beq
v_t=c\frac{\d \g_\pm}{\g_\pm}.
\eeq
For $\g_\pm=10^3,$ we get $v_t=3\times 10^7\d\g_\pm$~cm~s$^{-1}.$
\par
If we separate equation (27) into real and imaginary parts, we get two coupled simultaneous
equations. By solving them numerically, we find the growth rate $\G$ of stimulated Raman scattering.
Figure 2 shows the growth rate as function of $r_8$ and $\o_i/\o_p.$
Since the plasma density decreases with r as $1/r^3,$ growth rate decreases as $r_8$
increases. Also, if the radiation frequency becomes higher than the plasma frequency,
then radiation and plasma do not couple resonantly, which
leads to the decrease of growth rate at higher values of $\o_i/\o_p.$
\par
\vfill\eject
\hskip 30 cm .
\vskip -2 cm
\epsfysize12. truecm\hskip 2. truecm {\epsffile[18 176 558 716]{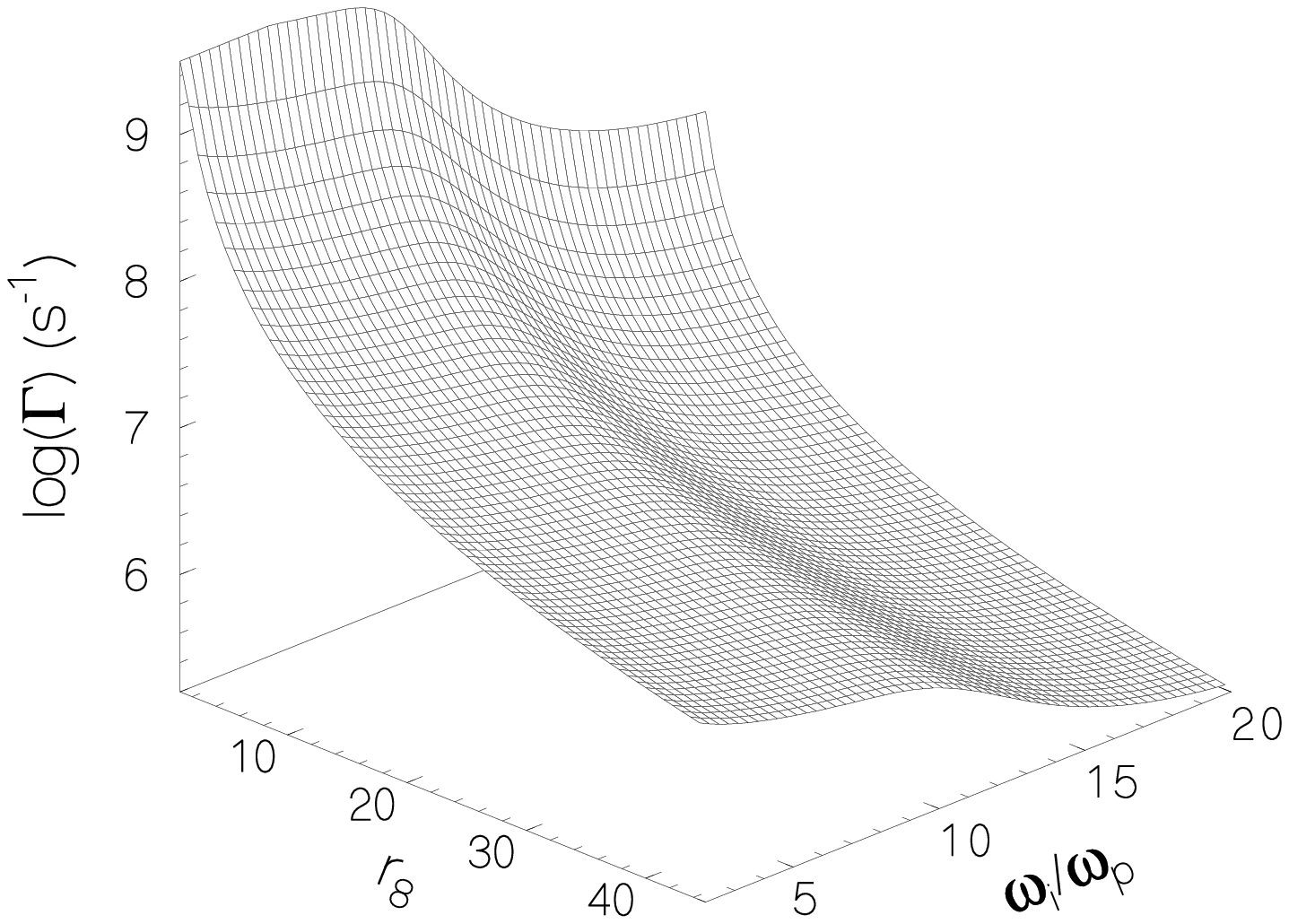}}
\vskip -2 cm
\noindent
{\small\rm
{\bf Figure~2.} The growth rate $\G$ of stimulated Raman scattering is plotted with
respect to $r_8$ and $\o_i/\o_p.$}
\vskip 0.5 cm
 To study the variation of $\G$ in the $\a_i$--$\d \g_\pm$ plane, we have made a contour
plot (Fig.~3), the labels on the contours indicate the values of log$(\G).$ Debye
length increases with the increase of $\d \g_\pm,$ therefore, plasma wave
experiences a large Landau damping, which leads to the drop in growth rate. The parameter
$\a_i$ is the ratio of amplitudes of electromagnetic waves, which are polarized in the
directions
parallel and perpendicular to the $\vec k_i-\vec B$ plane. When $\a_i$ is small, the
wave polarized in the direction parallel to $\vec k_i-\vec B$ plane becomes
strong and will have a
component along $\vec B.$  Therefore, the coupling between radiation and plasma
will be strong, which leads to the higher growth rate at the smaller values of $\a_i.$
\par
\epsfysize11.0 truecm\hskip 2cm {\epsffile[18 176 558 716]{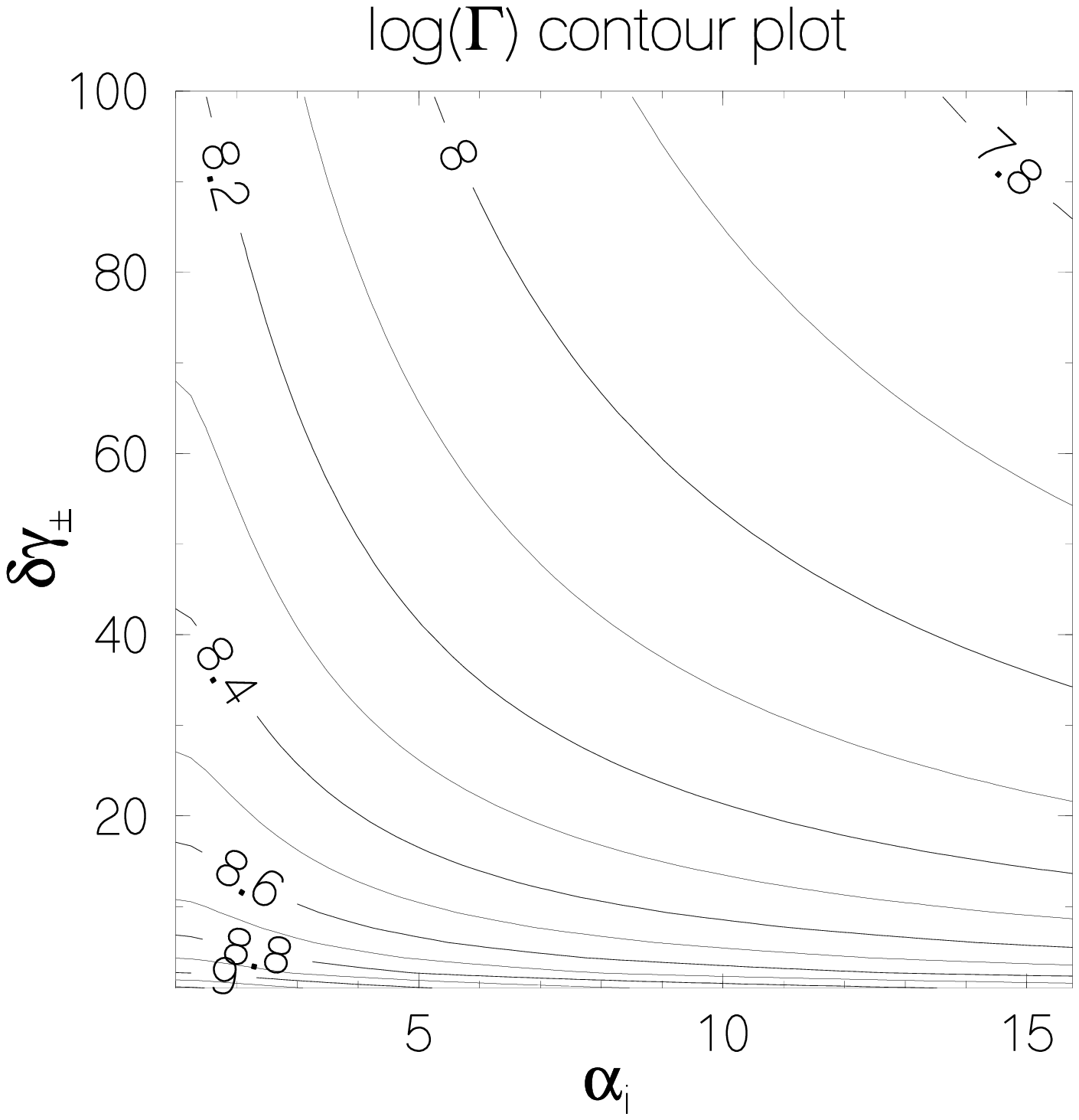}}
\vskip -1.5 cm
\noindent
{\small\rm
{\bf Figure~3.} Contour plot of $\G$ in the $\d \g_\pm$--$\a_i$ plane. The labels on the
  contours indicate the log($\G$) values.}
\vskip 0.5 cm
\par

For $\o\ll c^2\vec k_i\cdot\vec k/\o_i,$
Stokes mode becomes resonant, and anti-Stokes mode becomes non--resonant.
Then the equation~(27) can be written as
\beq
(\o-\o_l+{\rm i}\G_l)(\o-\o_l+{\rm i}\G_s)(\eta+\mu_a)=-10^{18}\frac{L_{30}}
{r_8^{7/2}\nu_{10}} \frac{(\mu_a+\a_i^2 \eta)}{(1+\a_i^2)} \quad {\rm rad^2}\,\,
{\rm s}^{-2},
\eeq
where $\o_l^2 = \o_{p}^2+(3/2)k^2v_t^2,$ and
\beq\G_l=\frac{\sqrt{\pi}}{2}\frac{\o_{p}}{(k\l_{D})^3}\exp\!\bc
[ -\frac{1}{2(k\l_{D})^2}-\frac{3}{2}\bc ]+\nu_c
\eeq
is the damping rate of the plasma wave, and the Debye length
$\l_{D}=v_t/\sqrt{2}\o_{p}.$ For $k\l_{D}\sim 1,$ we find
$\G_l\approx 2.5\times 10^9/r_8^{3/2}$~s$^{-1}.$
The pair plasma
collision frequency $\nu_c\approx 2.5\times 10^{-3} \ln\Lambda/r_8^3
\d\g_\pm^3$~s$^{-1}$ and Coulomb logarithm
$\ln\Lambda\approx 10.$ The collisional damping rate of the scattered
electromagnetic wave
is given by $\G_s=\o_{p}^2\nu_c/2\o_s^2\approx 0.06/\nu_{10}^2 r_8^6
\d\g_\pm^3$~s$^{-1}.$
\par
For $k\l_D > 1,$ plasma waves are highly damped and hence plasma losses its collective
behaviour.
Therefore, stimulated Raman scattering changes into the induced Compton scattering of electromagnetic
waves by the plasma particles. In the conventional treatment of induced Compton
scattering in pulsars (e.g., Blandford \& Scharlemann 1976; Sincell \& Krolik 1992) the
collective effects of the plasma are ignored. The collective treatment of the wave
scattering by plasma particles is justified if the condition $k \lambda_D \ll 1 $
or $v_{\rm ph}\gg v_t$ is met, where $v_{\rm ph}=w_p/k$ is the phase velocity of the
plasma wave.
This condition implies that the wave number of oscillation of the electrons in the
beat wave of the incident and scattered waves be much less than the inverse of the
Debye length.
It is not satisfied, when the beat wave is strongly Landu damped ($k \lambda_D > 1$) or
the beat wave will not feel the presence of a medium ($k \lambda_D \gg 1$), so that
the scattering process will be described as induced Compton scattering.
\par
 When $k\l_D \ll 1,$
by setting $\o=\o_{l}+{\rm i}\G, $ we can solve equation~(32) for
the growth rate:
\beq
\G=-{1\over 2}(\G_l+\G_s)\pm{1\over 2}
\sqrt{(\G_l-\G_s)^2+4.3\times 10^{18}\frac{L_{30}}
{r_8^{7/2}\nu_{10}}
\frac{(\mu_a+\a_i^2 \eta)}{(1+\a_i^2)(\eta+\mu_a)}}.
\eeq
Stimulated Raman scattering is a threshold process: if the intensity of the pump
exceeds the threshold, then only it would start converting its energy into
the decay waves. The threshold condition for the excitation of stimulated Raman
scattering is given by
\beq
\bb(\frac {L_{30}}{r_8^2}\bb)_{\rm thr}= 9.3 \times 10^{-19} \nu_{10}r_8^{3/2}
\G_l\G_s\frac{(1+\a_i^2)(\eta+\mu_a)}{(\mu_a+\a_i^2 \eta)}.
\eeq
The typical threshold intensities for stimulated Raman scattering are of the order
 of the observed intensities, implying that pulsar magnetosphere
may be optically thick to Raman scattering of electromagnetic waves.
\par
The growth rate just above the threshold is given by
\beq
\G=4.3\times 10^{8}\frac{L_{30}}{r_8^{2}\nu_{10}}
\frac{(\mu_a+\a_i^2 \eta)}{(1+\a_i^2)(\eta+\mu_a)}\quad{\rm s}^{-1},
\eeq
which is proportional to $L_{30}.$
The maximum growth rate attainable
for $\o_{p} > \G >\G_l,$ on the other hand, is
\beq
\G= 10^9\sqrt{\frac{L_{30}}{\nu_{10}^{3}r_8^{13/2}}
\frac{(\mu_a+\a_i^2 \eta)}{(1+\a_i^2)(\eta+\mu_a)}}\quad {\rm s}^{-1} .
\eeq

\section{Stokes parameters}

      When the phase matching conditions are satisfied, the growth rate $\Gamma$ becomes
large, and the scattered mode is amplified and become a normal mode of the plasma.
Under some conditions, the scattered mode leaves the plasma with polarization
which may be different from the polarization of the incident wave. The
polarization states of the
incident and scattered waves can be described more accurately using the
Stokes parameters (Rybicki \& Lightman~1979):
\beq
I_j= \eps_{xj}\eps_{xj}^*+\eps_{yj}\eps_{yj}^*,
\eeq
\beq
Q_j= \eps_{xj}\eps_{xj}^*-\eps_{yj}\eps_{yj}^*,
\eeq
\beq
U_j= 2\eps_{xj}\eps_{yj}^*\cos(\d_j)
\eeq
and
\beq
V_j= 2\eps_{xj}\eps_{yj}^*\sin(\d_j),
\eeq
where $j=i$ for the incident wave, and $s$ for the scattered wave.
The linear polarization is given by
\beq
L=\sqrt{U_j^2+Q_j^2},
\eeq
and the polarization position angle is given by
\beq
\chi_j=\frac{1}{2}\arctan({U_j}/{Q_j}).
\eeq
\par
 The transfer of energy between the modes will be efficient only when the
energy of the pump wave is strong enough to overcome the
damping losses or escape  of the generated waves.
Using the Manley--Rowe relation (Weiland \& Wilhelmsson 1977)
 \beq
{I_i\over \o_i}={I_s\over \o_s},
\eeq
we find the relation between
the incident flux $I_i$ and the scattered flux $I_s$:
\beq
I_s=\bb (1-{\o\over\o_i}\bb )I_i.
\eeq
   In the following two subsections, we consider the cases where the
polarization state of the incident wave is linear and circular,
and compute the polarization states of the scattered mode.

\subsection{Linearly polarized incident wave}
   Consider a linearly polarized electromagnetic wave ($\d_i=0,$ $\a_i=10$ and
$\chi_i=90^o),$ which excites stimulated Raman scattering in the magnetospheric plasma.
 Using the plasma and magnetic field parameters discussed
in the previous section, we computed the growth rate of stimulated Raman scattering. For
$k\l_D\ll 1,$ the instability becomes quite strong and the Landau damping of
Langmuir wave is minimal. Hence, the stimulated Raman scattering is resonantly
excited. Figure 4 shows the behaviour of linear (solid line) and
circular (broken line) polarization of scattered mode with respect to
$\eta$. It shows at some values of $\eta,$ which are close to 0.1, the linear
polarization of the incident wave can be converted almost completely into circular polarization of
scattered wave. Charge density variations within the pulse window could enhance the
conversion efficiency of linear to circular polarization to vary.
\par
\vskip -1.0 cm
\epsfysize10. truecm\hskip 2. truecm {\epsffile[18 176 558 716]{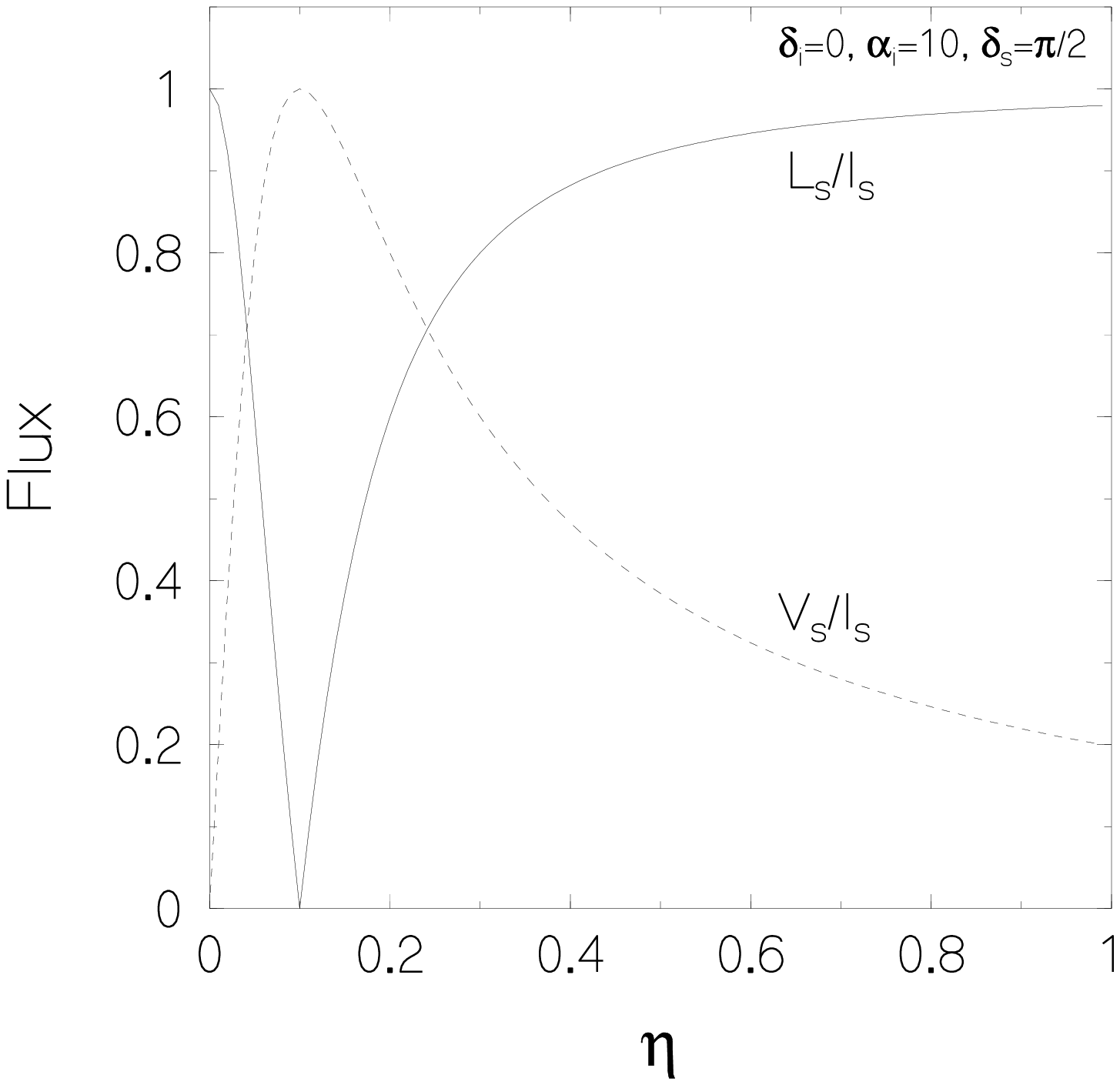}}
\vskip -0.5 cm
\noindent
{\small\rm
{\bf Figure~4.} The solid line and broken line curves indicate the variation of linear
($L_s$) and circular ($V_s$) polarization  of the scattered mode with respect to
$\eta.$ The normalizing parameter $I_s$ is the intensity of scattered mode.}
\vskip 0.5 cm
   The variation of polarization angle of the scattered wave with respect to $\eta,$
at different values of $\d,$ is shown in Fig.~5. For $\d=90^o$ and $\eta\leq 0.1,$ the
scattered mode becomes orthogonally polarized with respect to the incident wave.
Furthermore, if there is any variation in the plasma density or radiation intensity,
the value
of $\eta$ fluctuates and the scattered modes produced with $\eta\leq 0.1$ become
orthogonally polarized with respect to those produced with $\eta > 0.1.$
\vskip -1.5 cm
\epsfysize10.5 truecm\hskip 2cm {\epsffile[18 176 558 716]{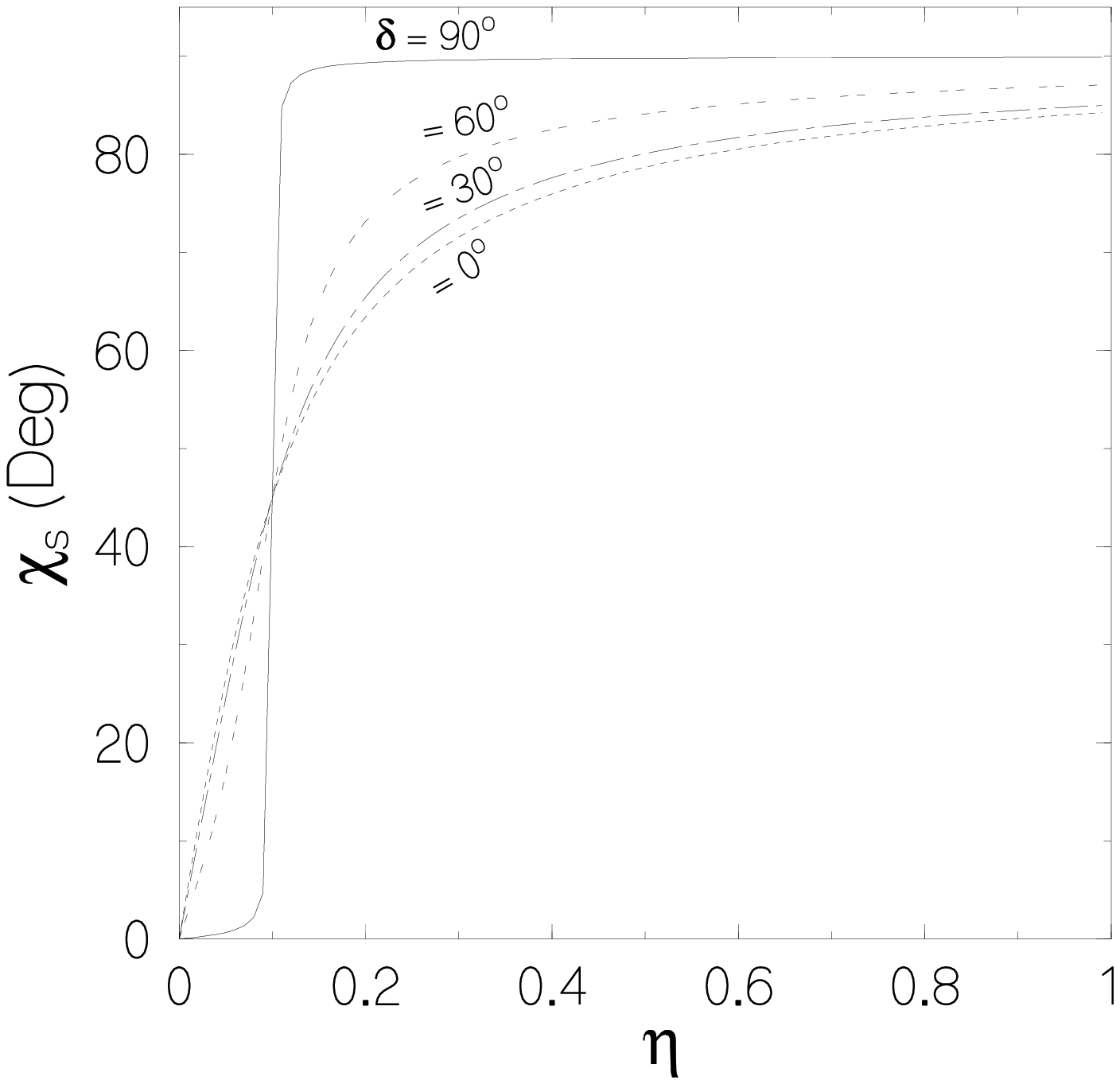}}
\vskip -1. cm
\noindent{\small\rm
{\bf Figure~5.} Polarization angle $\chi_s$ of the scattered mode is plotted as a function
of $\eta,$ at different values of $\d $ $(0^o,\, 30^o,\, 60^o$ \& $90^o).$ }

\subsection{Circularly polarized incident wave}

Suppose that the incident wave is circularly polarized ($\d_i=\pi/2$ and
$\a_i=1) $ then the scattered mode will be linearly polarized for $\eta\leq 0.2,$
while at the other values, both linear and circular polarizations with
different proportion exist, as indicated by Fig.~6. For
$\d_i< 0 $ and $ \d_s=\d_i-\d <0$ the sense
of circular polarization of the scattered mode becomes opposite to that of the
incident wave. Hence depending upon the plasma and radiation conditions,
stimulated Raman scattering can change the polarization of the pulsar radio signals.

\vskip -1. cm
\epsfysize10.5 truecm\hskip 2cm {\epsffile[18 176 558 716]{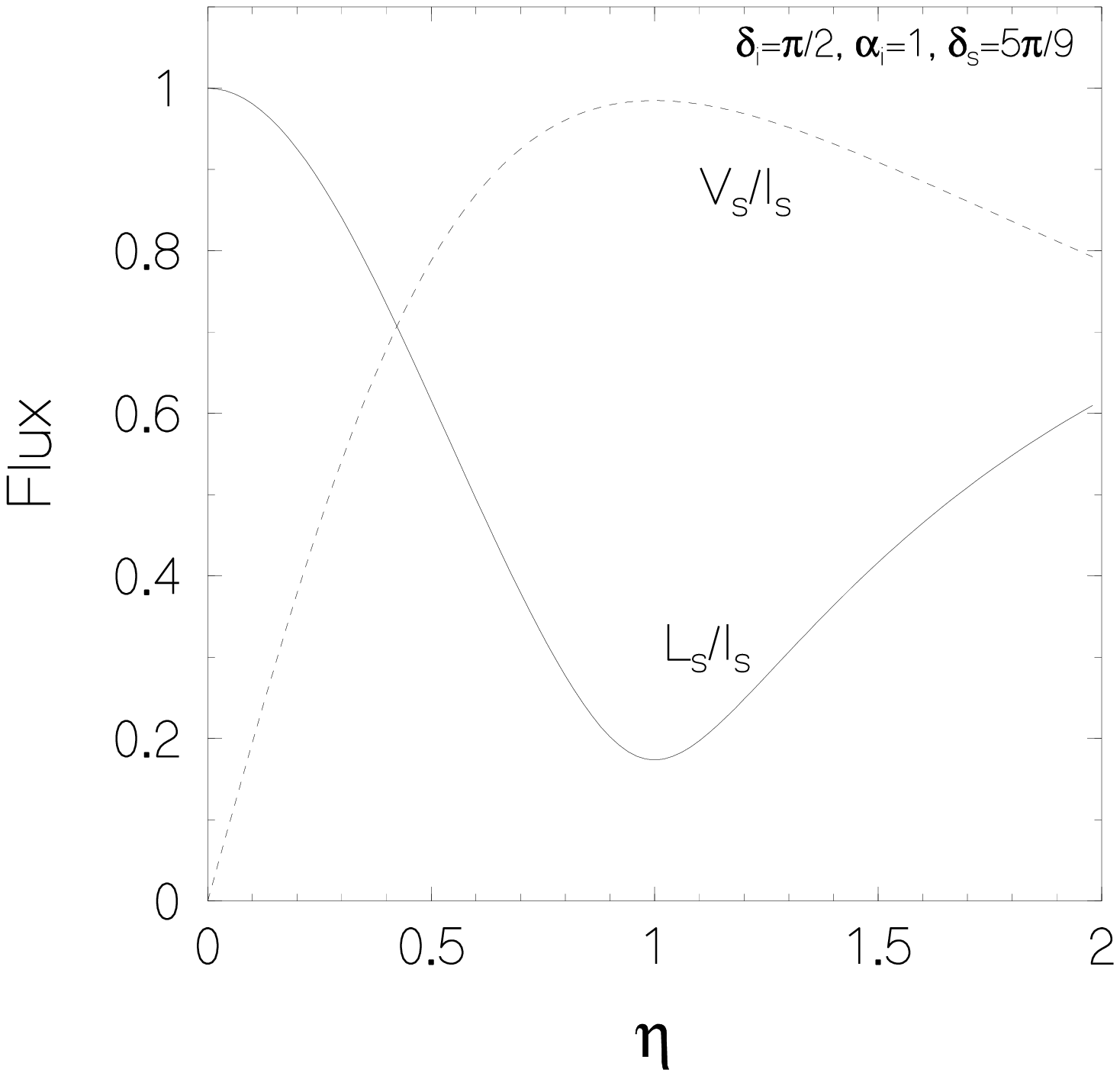}}
\vskip -0.5 cm
\noindent{\small\rm
{\bf Figure~6.}
The variation of linear ($L_s$) and circular ($V_s$) polarization  of the
scattered mode, with respect to $\eta,$ are indicated by the solid line and broken
line curves. The normalizing parameter $I_s$ is the intensity of the scattered mode.}

\section{DISCUSSION}

   The variable nature of circular polarization is evident from the polarization
distributions by Manchester, Taylor and Huguenin (1975), Backer and Rankin (1980)
and Stinebring et~al. (1984a,b). Very high degrees of circular polarization are
occasionally observed in individual pulses, even up to 100 per cent (Cognard
et~al. 1996), while the integrated or average pulse profiles
generally indicate much smaller degree of circular polarization (e.g., Lyne, Smith
\& Graham 1971), which show that the sign of circular polarization is variable
at any given pulse phase. Radhakrishnan and Rankin (1990) have identified
two extreme types of circular polarization in the observations: (a) an
antisymmetric type wherein the circular polarization changes sense in midpulse,
and (b) a symmetric type wherein it is predominantly of one sense.
The correlation of sense of antisymmetric type of circular polarization with
the polarization angle swing indicates the geometric property of emission process,
and is highly suggestive of curvature radiation.
\par
    The diverse nature of circular polarization may be
the consequence of pulsar emission mechanism and the subsequent propagation
effects in the pulsar magnetosphere (e.g., Melrose 1995; von~Hoensbroech, 
Lesch and Kunzl 1998). It seems rather difficult
to explain the various circular polarization behaviours using the widely
accepted magnetic pole models (e.g., Radhakrishnan \& Cooke 1969; Komesaroff 1970;
Sturrock 1971; Ruderman \& Sutherland 1975).
\par
The symmetric type of circular polarization observed in some pulsars
(e.g., PSR B1914+13, 0628-28) as shown in Figs.~7 and 8, may be due to the
propagation effects. However, it seems difficult for propagations effects
to explain, how the sign of the circular polarization can change precisely
at the center of the pulse in the case of antisymmetric type, as seen in
many pulsars e.g., PSR B1859+03 and B1933+16 (Rankin, Stinebring \& Weisberg 1989).
\vskip 0.5 cm
\epsfysize10.0 truecm\hskip 2cm {\epsffile[134 168 478 623]{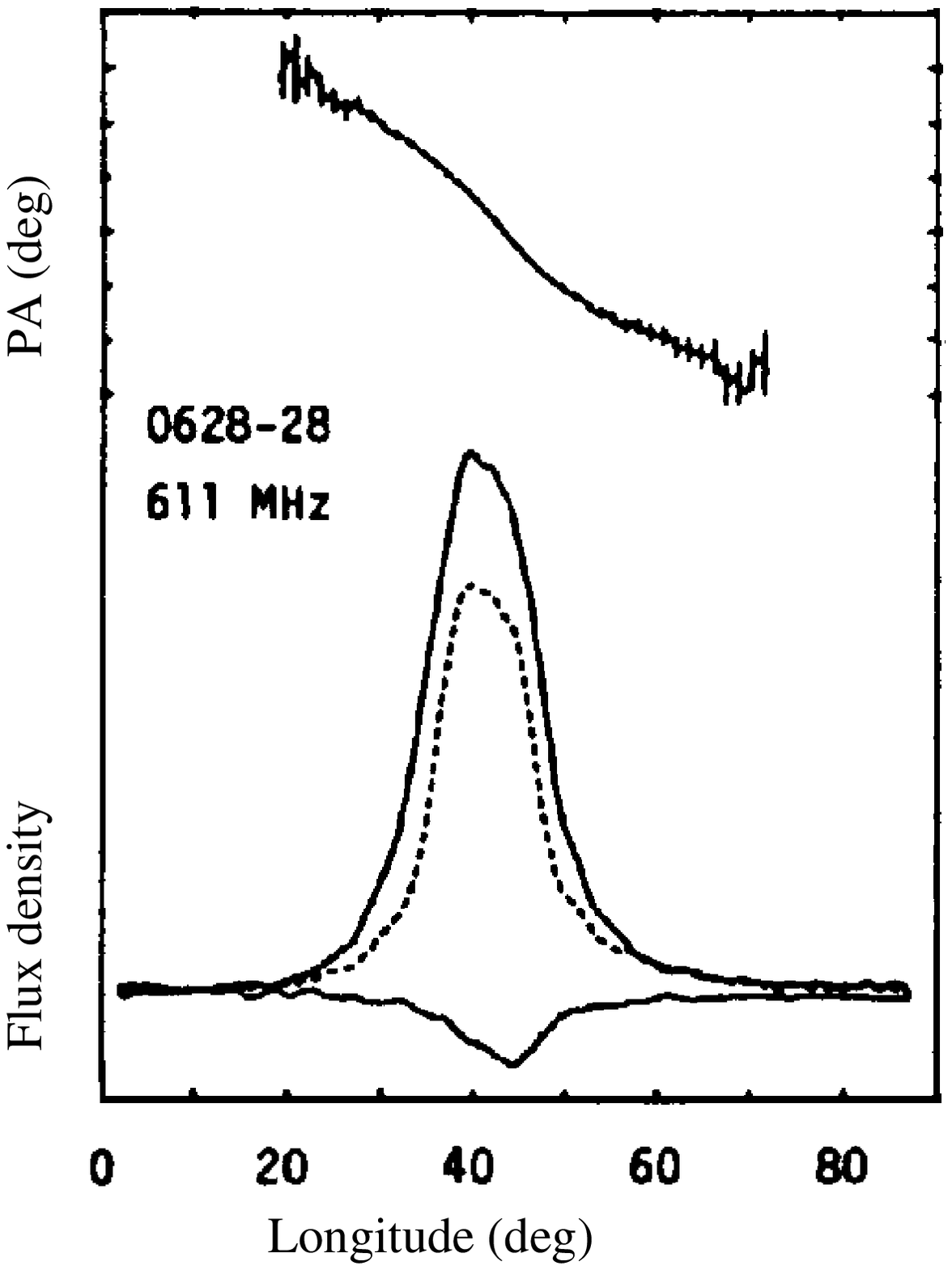}}
\vskip 0.3 cm
\noindent{\small\rm
{\bf Figure~7.}
PSR B0628-28, an example of a pulsar with `symmetric' circular polarization
 and high linear polarization (From Lyne \& Manchester 1988).}
\vskip 0.5 cm
   If we are to understand the radio emission mechanism, we must understand
the physical state of the radio-loud plasma in the polar cap. It is this plasma
that is the site of instabilities which are thought to produce coherent radio
emission. The role of different propagation effects on the pulsar polarization
has been discussed by Cheng and Ruderman (1979), Beskin, Gurevich and Istomin (1988)
and Istomin (1992). The mechanisms proposed by these authors predict a frequency
dependence for circular polarization, with weaker polarization at higher frequencies.
This is seen in some pulsars (e.g., PSR B0835-41, 1749-28, 1240-64) but
it is not generally the case (Han et~al. 1998). Istomin suggested that the linearly 
polarized incident wave becomes circularly polarized as a result of generalized 
Faraday rotation, however, it is observationally known that no generalized Faraday
rotation is evident in pulsar magnetospheres
(Cordes 1983; Lyne \& Smith 1990).
\vfill\eject
\hskip 30 cm .
\vskip -1.0 truecm
\epsfysize11.3 truecm\hskip 2cm {\epsffile[18 176 558 716]{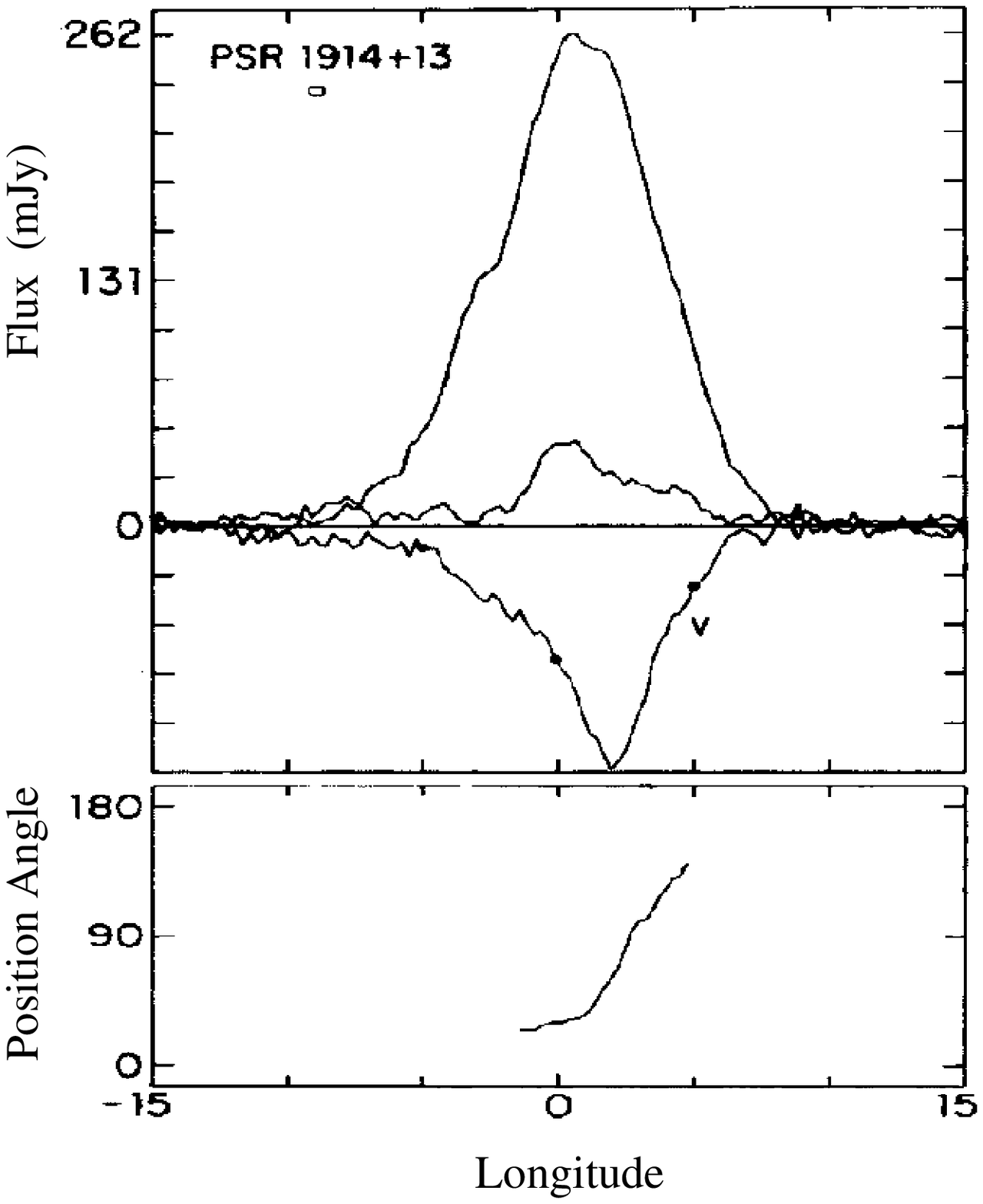}}
\vskip 0.5 cm
\noindent{\small\rm
{\bf Figure~8.}
Polarization of PSR B1914+13, a pulsar with strong circular polarization
over the whole observed profile (From Rankin, Stinebring \& Weisberg 1989).}
\vskip 0.5 cm
    We have presented a model to explain the polarization changes due to the
propagation of radio waves through the magnetospheric plasma. The features like
a large change in polarization angle, sense reversal of circular polarization and 
extremely rapid temporal changes in intensity would help us to
explain many observations, for which, the existing mechanisms proved to be
inadequate. Because of the very strong dependence of polarization angle on
plasma parameters via the growth rate, in an inhomogeneous plasma medium
the depolarization is a natural  outcome. We believe that the plasma process such
as the stimulated Raman scattering may be a potential
mechanism for the polarization variability in pulsars. The circular
polarization of a number of pulsars varies with frequency. The two clear
examples are PSR~B1240-64 and 2048-78, from which the opposite senses or
transitions of circular polarization have been observed at lower and higher
frequencies (Han et~al. 1998).

\section{CONCLUSION}

We considered the stimulated Raman scattering of the transverse electromagnetic
waves in the electron-positron plasma of pulsar magnetosphere.
The value of radiation--plasma coupling
parameter $\eta$
is determined by the polarization of the incident wave, and its value
can be determined only by the nonlinear analysis. In the process of three--wave
interaction, the phase matching condition (equation 13) between the initial phases
$(\d_i,$ $\d_s,$ $\d)$ and the value of $\eta$ determine the polarization state
of the scattered wave.
\par
 Many short time scale polarization variabilities, in individual pulses, such as
variations in amount of linear and circular polarization,
sense reversal of circular polarization and polarization angle swings
can be accounted for by considering the stimulated Raman scattering. The time scales
over which the changes takes place, 
in individual pulses, are of the order of e-folding time, which
is the inverse of the growth rate of stimulated Raman scattering.
\par
  It seems rather unlikely that the diverse behaviour of circular polarization can be
accounted for by a single mechanism. Both intrinsic emission and propagation effects
seem possible. The strong symmetric type of circular polarization observed in some pulsars is most
probably generated by propagation effects, such as 
the stimulated Raman scattering. Further simultaneous observations over a wide frequency
range would be valuable in sorting out the importance of propagation effects.

\section*{Acknowledgments}

We are  grateful to Y. Gupta and A. von~Hoensbroech for discussions and 
comments.
\vskip 0.2 cm
\begref
\bref Ass\'eo E. 1993, MNRAS 264, 940
\bref Ass\'eo E., Pelleiter G., Sol H. 1990, MNRAS 247, 529
\bref Backer D.C., Rankin J.M. 1980, ApJS 42, 143
\bref Barnard J.J., Arons J. 1986, ApJ 302, 138
\bref Beal J.H. 1990, in  Physical Processes in Hot Cosmic Plasmas, Eds. Brinkmann, W.,
        et al., (Kluwer Academic publications), 341
\bref Benford G. 1992, ApJ 391, L59
\bref Beskin V.S., Gurevich A.V., Istomin Ya.N. 1988, Ap. Space Sci. 146, 205
\bref Blandford R.D. 1975, MNRAS 170, 551
\bref Blandford R.D., Scharlemann E.T. 1976, MNRAS 174, 59
\bref Cheng A.F., Ruderman M.A. 1979, ApJ 229, 348
\bref Cognard I., Shrauner J.A., Taylor J.H., Thorsett S.E. 1996, ApJ 457, L81
\bref Cordes J.M. 1979, Space Sci. Rev. 24, 567
\bref Cordes J.M. 1983, in Positron--Electron pairs in Astrophysics,
      ed. Burns M.L.,Harding A.K., Ramaty R., AIP Conf. Proc. 101, 98
\bref Drake J.F., et al. 1974, Phys. Fluids 17, 778
\bref Fried D.,  Conte S.D. 1961,  The Plasma Dispersion Function,
                (Academic Press, New York), 1
\bref Gangadhara R.T., Krishan V. 1992,  MNRAS  256, 111
\bref Gangadhara R.T., Krishan V.  1993, ApJ  415, 505
\bref Gangadhara R.T., Krishan V.  1995, ApJ 440, 116
\bref Gangadhara R.T., Krishan V., Shukla P.K. 1993, MNRAS 262, 151
\bref Han J.L., Manchester R.N., Xu R.X., Qiao G.J. 1998, MNRAS, (submitted).
\bref von~Hoensbroech A., Lesch H., Kunzl T., 1998, A\&A 336, 209
\bref Hasegawa A. 1978, Bell System Tech. J.  57, 3069
\bref Istomin Ya.N. 1992, in Hankins T.H., Rankin J.M., Gil J.A., eds, The
   magnetosphere structure and emission mechanism of radio pulsars,
   Proc. IAU Coll. 128, 375
\bref Komesaroff M.M. 1970, Nat. 225, 612
\bref Krishan V., Wiita P.J. 1990, MNRAS 246, 597
\bref Kruer W.L. 1988, The Physics  of Laser-Plasma  interactions,
            (Addison-Wesley, New York), 70
\bref Lesch H., Gil J.A., Shukla P.K. 1994, Space Sci. Rev. 68, 349
\bref Liu C.S.,  Kaw P.K. 1976, Advances in Plasma Phys.,
          ed. Simon, A., \& Thompson, W. B., (Interscience New York), 6, 83
\bref Lominadze Dzh.G., Machabeli G.Z., Usov V.V. 1983, ApSS 90, 19
\bref Lyne A.G., Manchester R.N., 1988, MNRAS 234, 477
\bref Lyne A.G., Smith F.G. 1990,  Pulsar Astronomy, Cambridge Univ. Press,
        Cambridge, 234
\bref Lyne A.G., Smith F.G., Graham D.A. 1971, MNRAS 153, 337
\bref Manchester R.N., Taylor J.H., Huguenin G.R. 1975, ApJ 196, 83
\bref McKinnon 1997, ApJ 475, 763
\bref Melrose D.B. 1979, Aust. J. Phys. 32, 61
\bref Melrose D.B. 1995, JAA 16, 137
\bref Melrose D.B., Stoneham 1977, Proc. Astron. Soc. Australia 3, 120
\bref Michel F.C. 1982, Rev. Mod. Phys. 54, 1
\bref Radhakrishnan V., Cooke D.J. 1969, Astrophys. Lett.  3, 225
\bref Tsytovich V.N., Shvartsburg A.B. 1966, Sov. Phys. JETP 22, 554
\bref Radhakrishnan V., Rankin J.M. 1990, ApJ 352, 258
\bref Rankin J.M., Campbell D.B., Backer D.C. 1974, ApJ 188, 608
\bref Rankin J.M., Stinebring D.R., Weisberg J.M. 1989, ApJ 346, 869
\bref Ruderman M.,  Sutherland P. 1975,  ApJ 196, 51
\bref Rybicki G.B., Lightman A.P. 1979, Radiative Processes in
        Astrophysics,  (A Wiley--Interscience Publication), 62
\bref Sincell M.W.,  Krolik J.H. 1992, ApJ 395, 553
\bref Stinebring D.R., et al. 1984a, ApJS 55, 247
\bref Stinebring D.R., et al. 1984b, ApJS 55, 279
\bref Sturrock P.A. 1971, ApJ 164, 529
\bref Weiland J., Wilhelmsson H. 1977, Coherent Non--Linear Interaction
        of Waves in Plasmas, (Pergamon press), 60
\end{document}